\documentclass[conference]{IEEEtran}

\usepackage{amsthm,amssymb,graphicx,multirow,amsmath,color,amsfonts}
\usepackage[update,prepend]{epstopdf}
\usepackage[noadjust]{cite}
\usepackage[latin1]{inputenc}
\usepackage{bbm} 
\usepackage{pdfpages}
\usepackage{tabulary}
\usepackage{multirow}
\usepackage{comment}
\usepackage{array}
\usepackage{tabularx}
\usepackage{hyperref} 

\usepackage{tikz}
\usepackage{textcomp}
\usepackage{lipsum}

\newcommand\copyrighttext{  
	This work has been submitted to the IEEE for possible publication. 
	Copyright may be transferred without notice, after which this version 
	may no longer be accessible.
	
}
\newcommand\copyrightnotice{%
	\begin{tikzpicture}[remember picture,overlay]
	\node[anchor=north,yshift=-7pt] at (current page.north) {\fbox{\parbox{\dimexpr\textwidth-\fboxsep-\fboxrule\relax}{\copyrighttext}}};
	\end{tikzpicture}%
}

\usepackage{amsmath,amssymb}
\usepackage{caption}

\IEEEoverridecommandlockouts

\let\svbibcite\bibcite
\def\bibcite#1#2{\svbibcite{#1}{#2}}
\makeatletter
\let\svbiblabel\@biblabel
\def\@biblabel#1{\svbiblabel{#1}}
\makeatother

\begin{document}\vspace{-0.2cm}
\title{Toward Smaller and Lower-Cost  5G Devices with Longer Battery Life:  An Overview of \\ 
	3GPP Release 17 RedCap
} \vspace{-0.3cm}   

\author{\IEEEauthorblockN{Sandeep Narayanan Kadan Veedu$^1$, Mohammad Mozaffari$^2$, Andreas Höglund$^1$,\\ 
		Emre A. Yavuz$^3$, Tuomas Tirronen$^4$, Johan Bergman$^3$, and Y.-P. Eric Wang$^2$
	}\vspace{-0.25cm}\\
	\IEEEauthorblockA{
	\small  $^1$Ericsson Research, 164 83 Stockholm, Sweden. Emails: \{sandeep.narayanan.kadan.veedu, andreas.hoglund\}@ericsson.com\\  
	\small  $^2$Ericsson Research, Santa Clara, CA 95054, USA. Emails:{\{mohammad.mozaffari, eric.yp.wang\}@ericsson.com}\\  
	\small  $^3$Ericsson Business Unit Networks, 164 80 Stockholm, Sweden. Emails: \{emre.yavuz, johan.bergman\}@ericsson.com\\
	\small  $^4$Ericsson Research, 02420 Jorvas, Finland. Email: tuomas.tirronen@ericsson.com
\vspace{-0.2cm}}
		
	}
\maketitle
\copyrightnotice
\begin{abstract}
	
	The fifth generation (5G) wireless technology is primarily developed to support three classes of use cases, namely, enhanced mobile broadband (eMBB), ultra-reliable and low-latency communication (URLLC), and massive machine-type communication (mMTC), with significantly different requirements in terms of data rate, latency, connection density and power consumption. Meanwhile, there are several key use cases, such as industrial wireless sensor networks, video surveillance, and wearables, whose requirements fall in-between those of eMBB, URLLC, and mMTC. In this regard, 5G can be further optimized to efficiently support such mid-range use cases. Therefore, in Release 17, the 3rd generation partnership project (3GPP) developed the essential features to support a new device type enabling reduced capability (RedCap) NR devices aiming at lower cost/complexity, smaller physical size, and longer battery life compared to regular 5G NR devices. In this paper, we provide a comprehensive overview of 3GPP Release 17 RedCap while describing newly introduced features, cost reduction and power saving gains, and performance and coexistence impacts. Moreover, we present key design guidelines, fundamental tradeoffs, and future outlook for RedCap evolution.

\end{abstract}

\section{Introduction}
The fifth generation (5G) mobile communication technology was envisioned to enable a wide range of use cases and services with heterogeneous sets of requirements, outlined as part of the International Mobile Telecommunications - 2020 (IMT-2020) framework by the International Telecommunication Union Radiocommunication sector (ITU-R). The three main classes of 5G use cases are enhanced mobile broadband (eMBB), ultra-reliable and low-latency communication (URLLC), and massive machine-type communication (mMTC). To meet the requirements of these use cases, the 3rd generation partnership project (3GPP) introduced the 5G NR radio access technology in Release 15. NR was further enhanced in Release 16 and continues to evolve in Release 17 and beyond. In Releases 15 and 16, the focus was mainly on eMBB and URLLC types of services. 3GPP introduced two technologies, LTE for MTC (LTE-M) and Narrowband Internet of Things (NB-IoT) already in Release 13 focusing on mMTC requirements \cite{1}.

There are several emerging ``mid-range" IoT use cases that may currently not be best served by the 5G NR or LTE-M/NB-IoT. Examples of these use cases include wearables, industrial wireless sensors, and video surveillance. Some of the RedCap use cases can already today be adequately served by low-end LTE UE categories (e.g., Cat-1) for which there are no corresponding NR device types. To efficiently serve these use cases whose requirements on data rate, device cost, device size, and device battery lifetime fall in-between those of eMBB, URLLC, and mMTC, 3GPP has specified support for reduced capability (RedCap) NR devices in Release 17 \cite{2}. As such, RedCap can inherit most of the key benefits of NR, such as support of very wide range of frequency bands, including millimeter wave bands, network energy efficiency due to ultra-lean design, forward-compatible and beam-based air interface, and the ability to connect to the 5G core network (5GC) \cite{3}. Furthermore, RedCap can provide better coexistence with NR eMBB and URLLC on an NR carrier than LTE-based solutions, as well as provide a smooth migration path from LTE to NR for use cases that are currently addressed by LTE-based solutions \cite{4}. 

This article aims to provide an overview of RedCap from 3GPP standardization point of view. First, we describe the RedCap use cases and their specific requirements. Then we describe cost reduction and power saving techniques that were standardized in Release 17 to fulfill these requirements. We also discuss network impacts, and more specifically coexistence, coverage, and capacity impacts due to the introduction of RedCap in an NR network. Finally, we conclude by pointing to some further RedCap enhancements that may be pursued in future 3GPP releases.

\begin{table*}[t]
	\centering 
\caption{\small The specific requirements for RedCap use cases.}
\label{Table1}
	\begin{tabular}{|c|c|c|c|} \hline 
		& Wearables                             & Industrial wireless sensors & Video surveillance              \\\hline
		Data rate                & 5-50 Mbps in DL and                   &  2 Mbps   & 2-4 Mbps for economic video     \\  
		(Reference bit rate)     & 2-5 Mbps in UL\textsuperscript{1}     &                             & 7.5-25 Mbps for high-end video  \\ \hline 
		Latency                  & -                                     &  100 ms                     &  500 ms                         \\ \hline 
		Availability/reliability & -                                     & 99.99\%                     & 99\%-99.9\%                     \\ \hline 
		Device battery lifetime  & At least few days and up to 1-2 weeks & At least few years          & -                               \\ \hline 
		Traffic pattern          & -                                     & UL heavy                    & UL heavy                        \\ \hline 
		Stationarity             & Non-stationary                        & Stationary                  & Stationary                      \\ \hline 
		\multicolumn{4}{|l|}{Note 1:  Peak bit rate for wearables can be up to 150 Mbps in DL and 50 Mbps in UL.} 
		\\ \hline                          
	\end{tabular}\vspace{0.2cm}
\end{table*}

%


%

 \section{Use Cases and Requirements}

The technical work on RedCap was initiated as a study item in 3GPP Release 17. During the study item phase, the three reference IoT uses cases identified for RedCap were: wearables (e.g., smart watches, medical monitoring devices, AR/VR googles, etc.), industrial wireless sensors (e.g., pressure sensors, motion sensors, accelerometers, actuators, etc.), and video surveillance (e.g., surveillance camera for smart cities, factories, industries, etc.). The specific requirements on data rate, latency, availability/reliability, and battery lifetime of these use cases are summarized in Table \ref{Table1} \cite{2}. Clearly, requirements on data rate and latency that are not as demanding as for NR eMBB/URLLC, and the required battery lifetime is not at long as for mMTC. From Table \ref{Table1}, it can also be noticed that the requirements are quite diverse for the different use cases. However, it is desirable to support all the three uses cases with a single RedCap device type to reduce market fragmentation and maximize the benefits of economies of scale. It is worth clarifying that the use cases in Table \ref{Table1} are representative use cases. In practice, RedCap supports other use cases with similar sets of requirements.

In addition to the use case specific requirements, RedCap devices, in general, are also designed to enable  lower cost/complexity and smaller device size than high-end eMBB and URLLC devices. However, RedCap is not intended for low-power wide-area (LPWA) use cases, which are currently addressed by LTE-M and NB-IoT. The target coverage requirement for RedCap is the same as that for Release 15/16 NR devices. The 3GPP standard should also enable deployment of RedCap in all frequency division duplex (FDD) and time division duplex (TDD) bands in frequency range 1 (FR1), which ranges from 0.4 to 7.1 GHz, and frequency range 2 (FR2), which ranges from 24.2 to 52.6 GHz.

\begin{figure}[!t]
	\begin{center}
		\includegraphics[width=8.5cm]{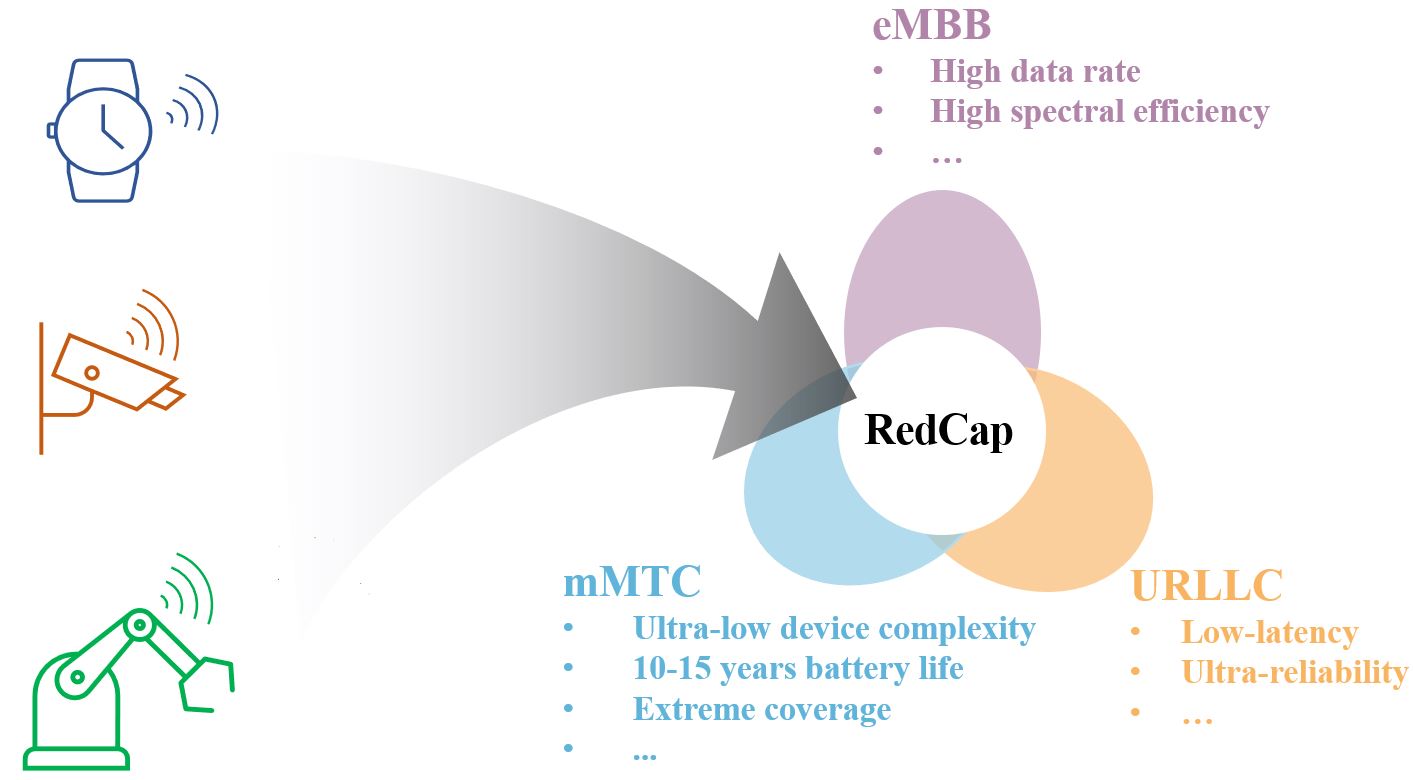}
		\caption{The requirements for RedCap in relation to that for eMBB, URLLC, and mMTC.}\vspace{-0.02cm}
		\label{RedCap}
	\end{center}
\end{figure}

Figure \ref{RedCap} provides an illustration of requirements for RedCap uses cases in relation to that for eMBB, URLLC, and mMTC use cases. 
The requirements discussed above have profound implications on the RedCap system design and standardization. During the study item phase, studies were carried out to identify technical solutions that can fulfil the requirements with minimal impacts to existing NR specifications. The technical solutions will be elaborated in the subsequent sections. The outcome of the study is documented in TR 38.875 \cite{5}. The study item phase was followed by a work item phase, resulting in the standardization of RedCap in Release 17 \cite{2}.

\begin{table*}[t]
	\centering 
	\caption{\small The comparison of the  reference Release 15 NR device and the simplest Release 17 RedCap device.}
	\label{Table2}
	\begin{tabular}{|p{3cm}|p{1.8cm}|p{1.8cm}|p{4cm}|p{4cm}|} \hline 
		\multirow{2}{*}{}   
        & \multicolumn{2}{l|}{\textbf{FR1 bands}}& \multicolumn{2}{l|}{\textbf{FR2 bands}}

         \\\hline                                                
		& Reference device       & RedCap device          & Reference device                & RedCap device                                                                                                                                                                                                                                                                                                                                                                          \\\hline
		Maximum device bandwidth                                 & 100 MHz                & 20 MHz                 & 200 MHz                                                             & 100 MHz                                                                                                                                                                                                                                                                                                                                                                                \\\hline
		Minimum antenna configuration$^{1}$ & 2 or 4 receiver branches                     & 1  receiver branch      &             2 antenna panels, each supporting 4 dual polarized antenna elements & 2 antenna panels, each supporting 2 dual polarized antenna elements
   \\\hline
		Minimum supported number of DL MIMO layers                         & 2 or 4                 & 1                      & 2                                                                   & 1                                                                                                                                                                                                                                                                                                                                                                                      \\\hline
		Maximum DL modulation order                              & 256QAM                 & 64QAM                  & 64QAM                                                               & 64QAM                                                                                                                                                                                                                                                                                                                                                                                  \\\hline
		Duplex operation                                         & TDD or full-duplex FDD & TDD or half-duplex FDD & TDD                                                                 & TDD                                                                                                                                                                                                                                                                                                                                                                                    \\\hline
		Cost reduction                                           & 0\%                    & $\sim$65\%  & 0\%                                                                 & $\sim$50\%                                                                                                                                                                                                                                                                                                                                                                  \\\hline
			\multicolumn{5}{|p{16cm}|}{$^{1}$In 3GPP the requirements on physical antenna implementation at the device are not specified for FR1 and FR2. However, the effective isotropic radiated power, the effective isotropic receiver sensitivity, and the spherical coverage requirements specified for different power classes in FR2 implicitly impose requirements on the actual antenna implementation. The minimum antenna configuration for the reference device indicated in this table assumes power class 3 \cite{6}. Both NR devices and RedCap devices may support a different power class with different set of requirements. The reference device and RedCap device support the same minimum number of receiver branches in FR2.}  \\\hline 
	\end{tabular} \vspace{0.25cm}
\end{table*}

\section{Reduced Device Capabilities}

The following cost/complexity reduction features were standardized as part of the Release 17 RedCap work item: 

\begin{itemize}
\item	Reduction of maximum device bandwidth.
\item	Reduction of minimum antenna configuration at the device.
\item	Reduction of minimum supported number of downlink (DL) MIMO layers.
\item	Relaxed maximum DL modulation order.
\item	Support for half-duplex operation in FDD bands.

\end{itemize}

The capabilities of a baseline RedCap device are compared with that of a reference NR device in Table  \ref{Table2}. For an objective comparison, the simplest Release 15 NR device is assumed as the reference. The set of reduced capabilities listed in Table \ref{Table2} can reduce the device modem bill-of-material cost by roughly 65\% for FR1 and 50\% for FR2.  More details on the evaluation methodology used for the cost reduction analysis and evaluation results can be found in \cite{5}, \cite{7}, \cite{8}.  Thus, the reduced capability features in Table \ref{Table2} can result in substantial cost reduction while at the same time performance, specification, and coexistence impacts are manageable. Reducing the device capabilities, in particular the antenna configuration, also contributes towards reducing the device size.

For a RedCap device with baseline (or minimum) capabilities as in Table \ref{Table2}, the achievable peak physical layer data rates are as follows:

\begin{itemize}
	\item	FR1 FD-FDD: $\sim$85 Mbps in DL and $\sim$90 Mbps in uplink (UL) assuming 15 kHz subcarrier spacing (SCS). For HD-FDD, the data rates will be lower as the device cannot transmit and receive at the same time. 
	\item	FR1 TDD: $\sim$60 Mbps in DL and $\sim$20 Mbps in UL assuming 30 kHz SCS and a DL/UL pattern of 3:1. 
	\item	FR2 TDD: $\sim$300 Mbps in DL and $\sim$100 Mbps in UL assuming 120 kHz SCS and a DL/UL pattern of 3:1.
	
\end{itemize}

 The above peak data rates are sufficient to fulfill the requirements of most of the intended uses cases for RedCap. The RedCap devices can also optionally support more advanced capabilities, such as up to 2 receiver branches and DL MIMO layers, 256QAM, and full-duplex FDD.  A RedCap device supporting these optional capabilities can achieve much higher peak data rates. However, a RedCap UE cannot support capabilities related to larger bandwidths than 20/100 MHz in FR1/FR2, carrier aggregation, dual connectivity, more than 2 device receiver/transmitter branches, or more than 2 DL/UL MIMO layers. These restrictions were made out of consideration for the device cost and due to the reason that these advanced capabilities are not needed to support intended use cases for RedCap. Other NR capabilities can be supported as optional features.

 In addition to the reduced capabilities in Table \ref{Table2}, which mainly concern the physical layer of the radio protocol stack, the following reduced capabilities for the higher layers were also introduced for RedCap devices:
 
 \begin{itemize}
 \item 	Reduction of maximum number of data radio bearers (DRBs) that the device must mandatorily support from 16 to 8. 
\item 	Reduction of sequence number (SN) length associated with the packet data unit for packet data convergence protocol and radio-link control  layers in acknowledged mode from 18 to 12 bits.
\item Support for automatic neighbor relation (ANR) functionality, which is a self-organizing network feature, is not mandatory. 
\end{itemize}  
It should be noted that the RedCap devices may optionally support 16 DRBs, 18-bit SN, or ANR. 

In addition to the aforementioned cost/complexity reduction features, there are a number of power saving features available for RedCap which will be discussed in the following section. \vspace{0.00cm}


 \section{Battery Lifetime Enhancement}
  The following power saving techniques that enable a longer battery lifetime were introduced (or enhanced) for RedCap  devices in Release 17.

\subsection{Extended DRX for RRC idle and inactive states}
  
 The data traffic between the device and the network can often be intermittent or bursty, i.e., there are periods of continual activity followed by long periods of inactivity for the arrival of data packets. The discontinuous reception (DRX) mechanism, which was introduced for NR in Release 15, allows the device to ``sleep" during these inactivity periods.  Specifically, in idle and inactive radio resource control (RRC) states, a device configured with DRX monitors the DL control channel (PDCCH) at paging occasion(s), based on the DRX cycle, to receive the scheduling information for a paging message. The device can enter a power saving state (e.g., turning off its receiver) during the remaining time, resulting in a significant reduction in the power consumption. In NR, the maximum value for the DRX cycle in the idle and the inactive states is 2.56 seconds.
 
\begin{figure}[!t]
  	\begin{center}
  		\includegraphics[width=8.3cm]{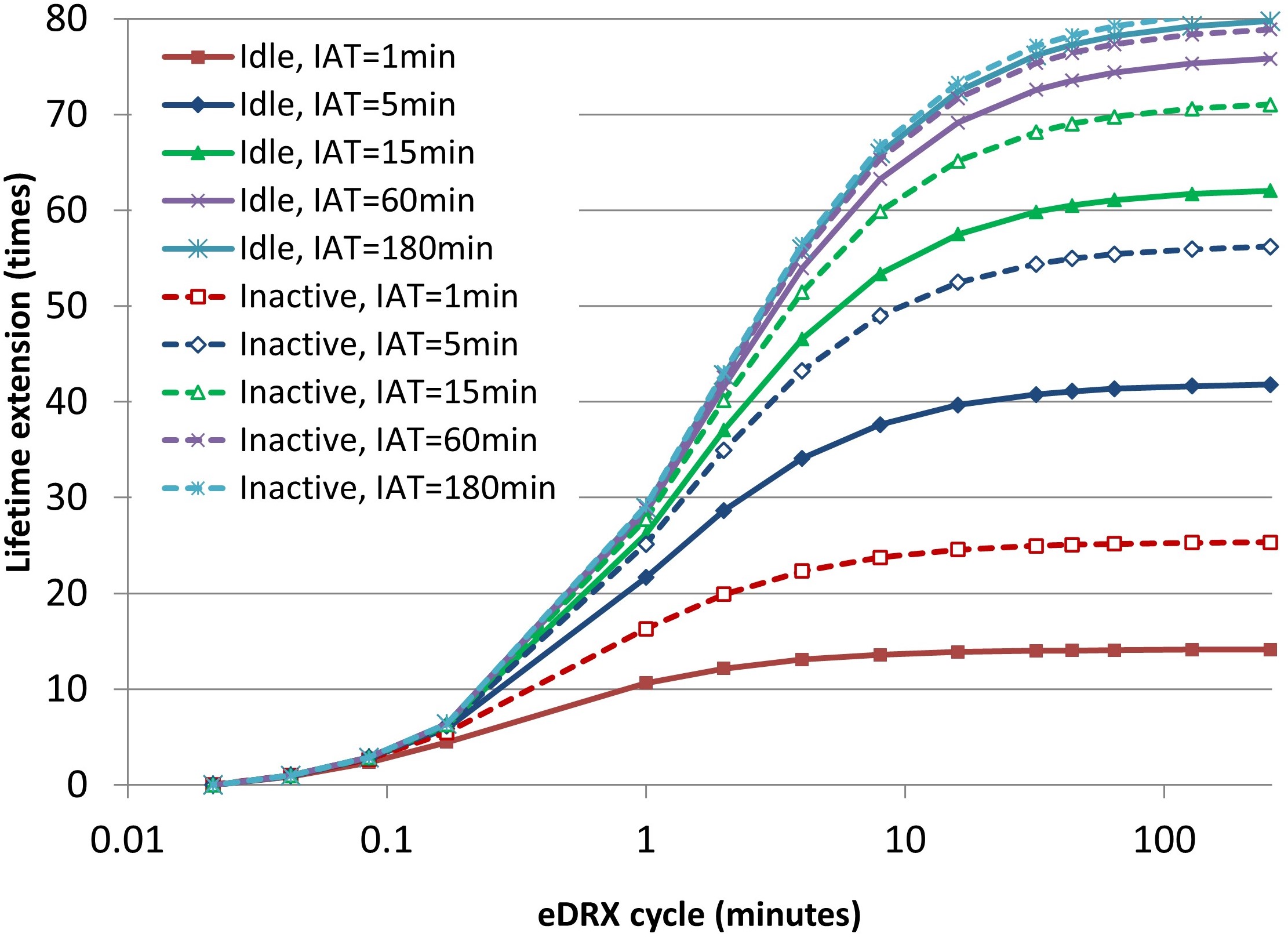}
  		\caption{The battery lifetime of a RedCap device in RRC idle and inactive states for different eDRX cycles and inter-arrival times.}\vspace{-0.1cm}
  		\label{Battery}
  	\end{center}
  \end{figure}

 With DRX, there is a tradeoff between power consumption at the device and DL latency (or DL reachability). However, RedCap use cases are, in general, not expected to have as tight or deterministic latency requirements as eMBB/URLLC use cases. Therefore, extended DRX (eDRX) with DRX cycles up to 10485.76 seconds (i.e., roughly 3 hours) in the RRC idle state and up to 10.24 seconds in the RRC inactive state were specified for RedCap in Release 17. For the RRC inactive state, where the device is in connected state from the 5GC perspective, eDRX cycles longer than 10.24 seconds are considered feasible from the radio-access network standpoint. However, 5GC  aspects needed further study, and therefore, introduction of the longer cycles in inactive state is postponed for consideration in Release 18 \cite{15}. The device battery lifetime extension for different eDRX cycles in the idle and the inactive states are shown in Figure \ref{Battery}. The results in Figure \ref{Battery} demonstrate that eDRX cycles of a few minutes can extend the battery lifetime by 10 to 70 times compared to the non-eDRX case, depending on the inter-arrival times (IAT) for data.  Thus, eDRX can help to meet the battery lifetime requirements for the most demanding RedCap use cases, such as industrial~wireless~sensors.
  
\subsection{RRM relaxation for stationary devices}

 In RRC idle and inactive states, the device needs to frequently perform radio resource management (RRM) measurements to ensure that the device is camping on the best available cell. The RRM measurements in the idle and inactive states are based on reference signal received power (RSRP) and reference signal received quality (RSRQ) values that are measured from the device's serving cell as well as its neighbor cells. Such measurements, although beneficial for achieving the best service quality once the device establishes a connection, will drain the battery even when there is no active data transmission between the device and the network. Therefore, already in Release 15, the device is allowed to skip RRM measurements for neighbor cells when the RSRP and the RSRQ of the serving cell are above certain thresholds, configured by the network.

In Release 16, additional relaxations for neighbor cell RRM measurements were introduced for low-mobility and not-at-cell-edge scenarios. More specifically, the device is allowed to relax neighbor cell measurements when the RSRP/RSRQ-based low-mobility condition is fulfilled for a period of time or when both low-mobility and not-at-cell-edge conditions are fulfilled. In Release 17, these conditions have been relaxed to allow for longer periods of relaxation for stationary RedCap devices. However, the potential additional gain on top of the existing Release 16 functionality is not expected to be significant. For more detailed results, see Annex E in \cite{5}. In Release 17, the network may furthermore configure the RSRP/RSRQ-based stationary condition for a device in RRC connected state and the device must report to the network when this condition is fulfilled or no longer fulfilled.

In addition to the eDRX enhancement and RRM relaxation features discussed above, the relaxation of PDCCH monitoring at the device by smaller numbers of blind decodes and control channel element limits was also studied during the RedCap study item phase.  However, this technique was not included in the scope of the subsequent Release 17 work item due to relatively small estimated potential power saving gains and significant performance impacts in terms of PDCCH blocking rate. In parallel to the work on Release 17 RedCap work item in 3GPP, power saving enhancements for ordinary types of NR devices were specified in a separate work item \cite{9}. These enhancements, which include paging early indication and dynamic search space set switching, as well as the power saving techniques that were standardized in Release 15 and Release 16, are available as optional features also for RedCap devices.

\section{Coexistence, Coverage, and Capacity Impacts}  
 
 Here, we discuss various system impacts and design considerations for RedCap. 
  
 \subsection{Coexistence with other NR devices} 
  
  Ensuring coexistence between RedCap and other NR devices was one of the key design objectives in the RedCap work item. As a result, RedCap devices are designed to efficiently coexist with other NR devices. However, some coexistence challenges were identified during the work on RedCap. In what follows, we describe two of the main ones and how the Release 17 specification permits resolving them. 
  
 \subsubsection{Identification of RedCap devices}
  Many of the reduced capabilities listed in 
 Table \ref{Table2} may impact how the devices are scheduled during the random-access procedure, in particular when RedCap devices and other NR devices are deployed on the same NR carrier. In NR, the information on devices' capabilities is in many cases known to the network only after the device establishes connection with the network, which occurs after completion of the random-access procedure. Until that point, the network schedules the device with only the minimum capabilities that are supported by all the devices. For a cell that supports both RedCap devices and other NR devices, the network may need to schedule all devices based on the minimum capabilities of the RedCap devices. To overcome this limitation and to allow for an efficient use of network and device resources, an indication is introduced in Release 17 to identify during the random-access procedure whether the device has reduced capabilities compared to legacy devices.  More specifically, the indication is provided in Message 3 (or Message A) of the random-access procedure \cite{10} in the form of a RedCap-specific logical channel ID value for Common Control Channel (CCCH). Additionally, if RedCap-specific physical random-access channel (PRACH) resources are configured in the cell, the indication is also provided implicitly already in Message 1.

  \subsubsection{Avoidance of PUSCH resource fragmentation}
  
  One of the basic features in NR is the so-called bandwidth part (BWP), which allows a device to operate with a narrower bandwidth than the cell carrier bandwidth. This also helps to reduce the device power consumption and enables support of devices with different bandwidth capabilities \cite{11}. In general, the device is not required to transmit or receive outside the frequency range spanned by its active BWP. The random-access procedure is carried out on the initial UL/DL BWP configured in the system information. The initial BWPs for regular NR devices can be much wider than the maximum bandwidth supported by a RedCap device (20/100 MHz in FR1/FR2). In fact, a typical network configuration is to have the UL BWP as wide as the available carrier bandwidth with the PUCCH transmission frequency hopping between the edges of the carrier. This helps to avoid fragmentation of the contiguous PUSCH resources and minimize the resulting reduction in UL peak data rate caused by PUCCH transmissions. Note that it is desired to avoid fragmenting the PUSCH resources since Release 15/16 NR devices are not typically expected to support non-contiguous PUSCH resource allocation. The initial DL BWP may be configured with any defined bandwidth up to the maximum device bandwidth. For Release 15/16 NR devices, it contains the entire frequency span of the synchronization signal and PBCH block (SSB) and the control resource set with ID = 0 (CORESET\#0), and in case of TDD its center frequency aligns with that of the UL BWP.

\begin{figure}[!t]
    	\begin{center}
    		\includegraphics[width=8.5cm]{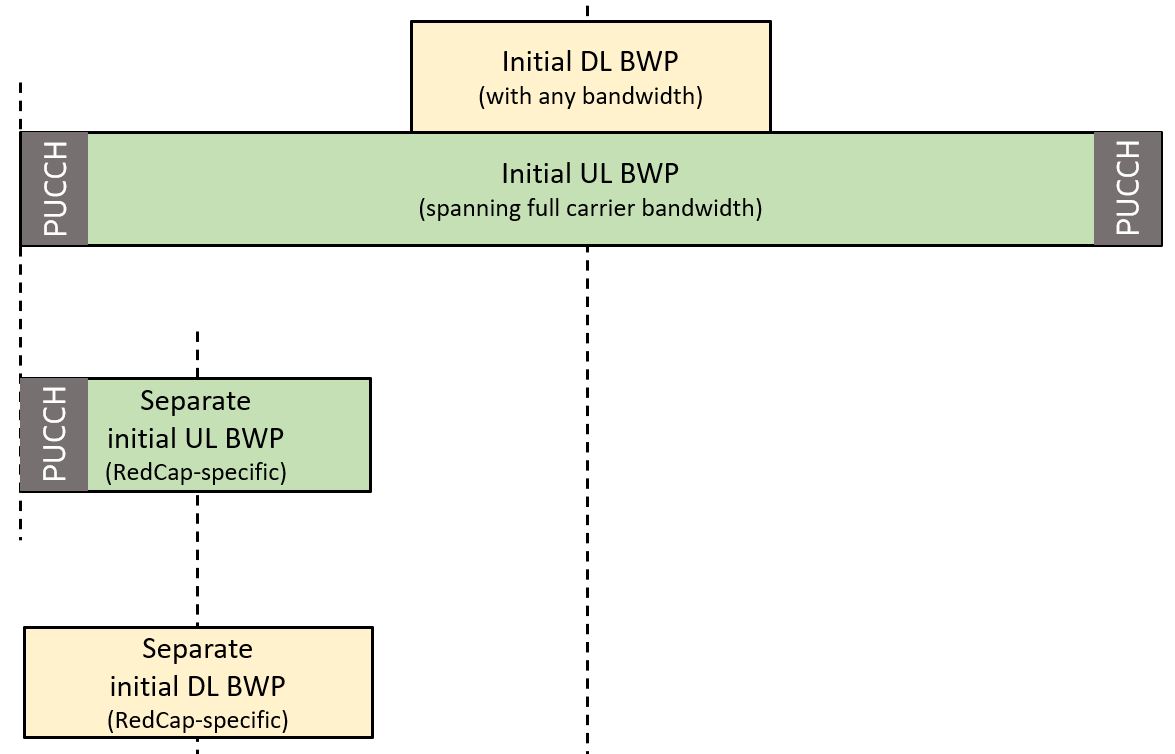}
    		\caption{ An example of network configuration when RedCap and other NR devices are deployed on the same NR carrier.}\vspace{-0.02cm}
    		\label{BWP}
    	\end{center}
    \end{figure}

  Since the initial BWPs configured for regular NR devices in the cell may be too wide for RedCap devices, the Release 17 specification supports configuration of separate RedCap-specific initial UL and DL BWPs. However, if the separate UL/DL BWP is placed near the middle of the carrier, PUCCH transmissions from the RedCap devices on the edges of this UL BWP might cause PUSCH resource fragmentation for other devices. This issue can become exacerbated by the fact that PUCCH frequency hopping cannot be disabled on an initial UL BWP in the pre-Release-17 specification. Configuration of the separate UL BWP near the carrier edge can minimize the fragmentation, but this may cause center frequency misalignment between the UL and the DL BWPs in TDD. Placement of the separate DL BWP near the carrier edge may not be possible as the SSB and the CORESET\#0 may be configured around the middle of carrier. Therefore, to alleviate the concerns of resource fragmentation, the Release 17 specification also supports disabling of PUCCH frequency hopping on the separate initial UL BWP for RedCap devices, as well as the potential configuration of the separate initial DL BWP without the presence of the SSB and the CORESET\#0. These enhancements enable the placement of the separate UL/DL BWPs near the carrier edge, and thereby avoid to a large extent the PUSCH resource fragmentation issue. In fact, it is important to have proper configurations for UL/DL BWPs during and after initial access to ensure an efficient coexistence between RedCap and regular NR devices. Figure \ref{BWP} presents an exemplary network configuration when RedCap and other NR devices are deployed on the same carrier. 
  
  Meanwhile, there are several tradeoffs and design considerations in terms of BWP configuration. Specifically, considering tradeoffs between configuration flexibility, UL data rate, and signaling overhead the following design aspects can be envisioned:
  
  \begin{itemize}
  	 
  	\item Depending on the carrier bandwidth, peak data rate requirements, device capability, and traffic activity the network can determine suitable configurations for RedCap (e.g., the location of UL/DL BWPs). 
  	\item From signaling overhead perspective, it is desired to have shared BWP configurations for RedCap and regular NR UEs. In particular, with a shared DL BWP between RedCap and regular NR devices, there is no need to provide additional BWP configurations and transmit additional SSBs which is beneficial in terms of signaling overhead, network energy efficiency, and inter-cell interference.
  	
  \end{itemize}

  In addition to the enhancements to resolve coexistence challenges described above, the specification also provides communication service providers the ability to bar RedCap devices from accessing the cell carrier using an indication in system information (SIB1). This may be useful in scenarios where the service providers may want to prioritize the available radio resources for serving other NR devices in the cell in the event of network overload, or there may be suspicion that allowing access to RedCap devices may impact the performance of the other devices.

  \subsection{Coverage impacts}
  
  The reduction of device capabilities described in Section III can enable both lower device cost and smaller device size, but they can also have detrimental impacts on the coverage of the devices. In particular, the reduction of antenna configuration and the reduction of device bandwidth have the largest impacts on the coverage. Therefore, evaluation of coverage impact was carried out as part of the study item. The coverage impact evaluation was based on link budget analysis. All the UL and the DL physical control and data channels, including those that carries the random-access messages, such as Message 1 (PRACH), Message 2 (PDSCH), Message 3 (PUSCH), and Message 4 (PDSCH), PUCCH, and the PDCCH scheduling Message 2/4 on common search space were included in the link budget analysis. The analysis considered different deployment scenarios, such as rural and urban (both macro and micro, with DL power spectral densities of 33 dBm/MHz and 24 dBm/MHz, respectively) in FR1 and indoor in FR2.

  The metric used in the link budget analysis is maximum isotropic loss (MIL), which captures path loss as well as beamforming gains at the base station and the device. The target coverage requirement for each of the channels of a RedCap device within a deployment scenario was deemed to be the MIL of the bottleneck channel of a reference device within the same deployment scenario. The bottleneck channel is the channel with the lowest MIL, and reference device is a Release 15/16 NR device with only the baseline capabilities. That is, if the MIL of any of the channels of a RedCap device is lower than the lowest MIL of the reference device, coverage recovery would be needed for the corresponding channel(s). The amount of recovery needed for a specific channel is the difference between the MIL of that channel and the MIL of the bottleneck channel of the reference device.

  The outcome of the link budget analysis can be summarized as follows: 
  
  \begin{itemize}
  	\item	In the UL in FR1 and FR2, coverage recovery may not be needed for any of channels. 
  	\item	In the DL in FR1, coverage recovery is needed for Message 2 for a RedCap device with 1 receiver branch in the urban micro scenario. Coverage recovery may not be needed for a RedCap device with 2 receiver branches, neither in the urban micro nor the urban macro scenario.  
  	\item	In the DL in FR2, the need for coverage recovery depends on the choice of maximum transmitted radiated power (TRP) for the RedCap and the reference devices.  If a TRP of 23 dBm is assumed, coverage recovery may be needed for some DL channels (assuming no reduction in the number of antenna elements), such as Message 2, Message 4, and PDSCH. If, on the other hand, a TRP of 12 dBm is assumed, coverage recovery may not be needed for the DL channels. 
  \end{itemize}
  	
 It is worth mentioning that the link budget analysis carried out as part of the study item assumed 3 dB worse transmitter/receiver antenna efficiency in FR1 for the RedCap device than the reference device \cite{5}. This was done to capture the impact of smaller RedCap device size. However, this is a conservative assumption, and hence, no such assumption was considered during the subsequent work item and in the outcome of the link budget analysis summarized above. For the coverage recovery of channels identified above, techniques specified in Release 15/16 (e.g., TBS scaling for Message 2, HARQ re-transmissions for Message 4 and PDSCH, etc.) can be used. In addition, if needed, coverage enhancements that were specified in a separate work item in Release 17 may also be used \cite{12}. Therefore, no new coverage recovery techniques were specified as part of the RedCap work item. More details on the coverage impacts and the evaluation results can be found in \cite{5} and \cite{13}.

 \subsection{Capacity impacts} 	
\begin{figure}[!t]
 	\begin{center}
 		\includegraphics[width=9cm]{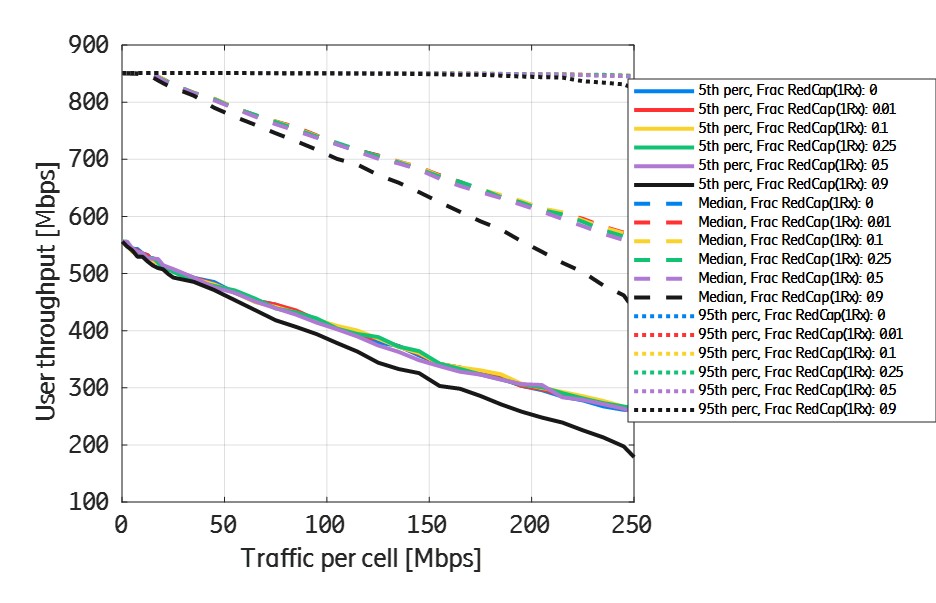}
 		\caption{eMBB downlink user throughput vs. cell load for increasing fraction of 1 receiver branch RedCap users (2.6 GHz).}\vspace{-0.02cm}
 		\label{Capacity}
 	\end{center}
 \end{figure}

  	It may be a concern that the lower capabilities of RedCap devices could have a negative impact on system performance, and e.g., lower throughput for eMBB users. To evaluate this, system level simulations were performed in line with the assumptions in \cite{5}. Figure \ref{Capacity} shows the DL user throughput for eMBB users as a function of the cell load for an increasing fraction (0\% to 90\%) of RedCap devices with 1 receive branch in FR1 (2.6 GHz frequency band).

  The UL throughput is, however, not impacted by the increasing fraction of RedCap devices and therefore not shown here. From Figure \ref{Capacity}, it is seen that for the eMBB users with good radio link quality (95th percentile, dotted curves) the DL user throughput is mainly unaffected by the amount of RedCap users.  For the median users (dashed curves) and the 5th percentile (full curves), it can be seen that the eMBB user throughput decreases somewhat with increasing load and increasing RedCap user fraction. This is due to the lower spectral efficiency of RedCap which results in a higher resource utilization at a given offered load. However, with the assumed traffic modeling from \cite{5}, an eMBB user will generate a load of $2\times 10^7$ bits/s (0.5 MB payload every 200 ms) and a RedCap user will generate a load of $4\times 10^5$ bits/s (0.1 MB payload every 2 s), which means the load from a RedCap user is 50 times lower than that from an eMBB user. Therefore, any differentiation in the throughput results is seen at first for a 90\% RedCap user fraction and having 90 times more RedCap devices than eMBB devices in a cell may not be a realistic scenario.

%

%

\section{Conclusion and Future Evolution}\vspace{-0.01cm}
To expand the 5G device ecosystem and to cater to a mid-range IoT market segment which may not yet be best served by the existing NR standard, 3GPP has introduced RedCap in Release 17. This article has provided a comprehensive overview of RedCap by focusing on how the 3GPP standard enables a device design that fulfills the requirements of the mid-range IoT use cases. As highlighted in this article, RedCap devices may have substantially lower cost, smaller size, and longer battery lifetime compared to other NR devices, while at the same time achieving higher data rates and lower latency than IoT devices based on LTE, LTE-M, or NB-IoT. The article also sheds light on how the standard allows for good coexistence between RedCap and other NR devices when they are deployed on the same 5G network. Furthermore, the article also describes the potential coverage and capacity impacts associated with the reduced device capabilities.

Further enhancements of RedCap are expected in 3GPP Release 18, branded as the first release of 5G Advanced, symbolizing a major evolution of 5G. The Release 18 enhancements for RedCap will be built on the foundation that had been laid in Release 17. The enhancements could provide improved support for Release 17 use cases, as well as support expansion into new segments of use cases, for e.g., smart grid. To address the new use cases, 3GPP has agreed to study further device simplifications \cite{14} and battery lifetime enhancements \cite{15}. It is, however, important to ensure that the integrity of the RedCap ecosystem is maintained in future releases so that the benefits of economy of scale can be maximized. 

\def\baselinestretch{1.00}
\bibliographystyle{IEEEtran}
\bibliography{referenceConf}
\vspace{0.9cm}

\end{document}